\documentclass[useAMS,usenatbib]{mn2e}

\title[Power-law S\'{e}rsic Profiles]
{Power-law S\'{e}rsic profiles in hydrostatic stellar galaxy discs}

\author[C. Struck and B. G. Elmegreen]
{Curtis Struck,\thanks{E-mail: curt@iastate.edu (CS);
bge@watson.ibm.com (BGE)}$^1$
Bruce G. Elmegreen$^{2}$ \\
$^1$ Dept. of Physics and Astronomy, Iowa State Univ., Ames, IA 50011 USA\\
$^2$ IBM Research Division, T.J. Watson Research Center, 1101 Kitchawan Road,
Yorktown Heights, NY 10598, USA}

\usepackage{amssymb}
\usepackage{amsmath}
\usepackage{graphicx}
\usepackage{multirow}
\def\aap{{ A\&A}}

\def\aj{{AJ}}

\def\apj{{ApJ}}
\def\apjl{{ApJL}}

\def\mnras{{MNRAS}}

\def\apjs{{ApJS}}
\def\araa{{ARA\&A}}

\begin{document}
\date{\today}

\pagerange{\pageref{firstpage}--\pageref{lastpage}} \pubyear{0000}

\maketitle

\label{firstpage}
\begin{abstract}
Previously, we showed that surface density profiles of the form of a power-law times a S\'{e}rsic function satisfy the hydrostatic Jeans equations, a variety of observational constraints, and the condition of a minimal radial entropy profile in two-dimensional galaxy discs with fixed power-law, halo potentials.  It was assumed that such density profiles are generated by star scattering by clumps, waves, or other inhomogeneities.  Here we generalize these models to self-gravitating discs. The cylindrically symmetric Poisson equation imposes strong constraints. Scattering processes favor smoothness, so the smoothest solutions, which minimize entropy gradients, are preferred. In the case of self-gravitating discs (e.g., inner discs), the gravity, surface density and radial velocity dispersion in these smoothest models are all {\it of the form $1/r$ times an exponential.} When vertical balance is included, the vertical velocity dispersion squared has the same form as the surface density, and the scale height is constant. In combined self-gravitating plus halo gravity cases, the radial dispersion has an additional power-law term. Nonetheless, the surface density profile has the same form at all radii, without breaks, satisfying the `disc-halo conspiracy'. The azimuthal velocity and velocity dispersions are smooth, though the former can have a distinct peak. In these models the vertical dispersion increases inwards, and scattering may mediate a transition to a secular bulge. If halo gravity dominates vertically in the outer disc, it flares. The models suggest a correlation between disc mass and radial scale length. The combination of smoothness, simplicity, ability to match generic observational features and physical constraints is unique to these models.
\end{abstract}


\section{Introduction}

For nearly fifty years, exponential fits to the surface brightness profiles of galaxy discs have been commonly used (\citealt{fr70}, \citealt{va02}). Whether this profile form is the result of physical processes or a fortuitous approximation has not always been clear, since early observations did not extend to very faint surface brightness levels. However, in recent years the observations have gone much deeper, confirming the exponential fit over 10 scale lengths in an increasing number of cases (e.g., \citealt{bl05}  \citealt{ga09}, \citealt{er05, er08}, \citealt{po06}, \citealt{he13}, and \citealt{zh15}, \citealt{hi16},  \citealt{tr16}, \citealt{wa16} \citealt{raj19}, \citealt{sta19}). The accuracy of the fit over such a large dynamic range suggests there is a physical cause.

In a series of recent papers we have explored the theory that the profile form in galaxy discs is the result of scattering of stars, which modifies any initial profile until a universal steady form is obtained. This theory is distinct from the radial migration theory of Sellwood and Binney (\citealt{se02}, also see \citealt{be15}), although both involve stellar scattering. \citet{se02} demonstrated that as spiral waves develop, the number of stars on horseshoe orbits in the vicinity of wave co-rotation grows, as does the region containing these orbits. The orbital modification leads to radial migration, but the process is mild with little change in the stellar angular momentum distribution, and orbits remain nearly circular. Alternately decaying and redeveloping waves, with different co-rotation radii spread the effects across the disc, and drive the evolution of the stellar profile. This process is not rapid, e.g., the recent work of \citet{da18} shows that relatively few particles are involved at any one time. \citet{be15} however, show that it smooths inhomogeneities due to accretion, but reforms an exponential profile slowly at best.

In \citet{el13} we considered a different limit, massive local scattering centers distributed across a galaxy disc, which produce impulsive scattering events rather than the continuous orbit modification over some duration as in the Sellwood and Binney model. The original motivation was to study the stellar profile evolution in young discs at high redshift whose morphology is dominated by massive clumps rather than spiral waves. The possibility that strong spiral waves or bars could also scatter impulsively rather than continuously, and that there might be 'mixed mode' scattering was also considered. The scattering models of \citet{el13} did indeed show that impulsive scattering produces exponential profiles in a wide range of gravitational, potentials. However, a significant fraction of the resulting orbits were quite eccentric, with high apparent velocity dispersions, more appropriate to a thick disc component.

Subsequent work showed that extended, or less concentrated scattering centers, like gas shells and holes rather than compact clumps, would produce fewer very eccentric orbits, while still producing exponential profiles (\citealt{st17b}). This work is particularly relevant to dwarf irregular galaxies, which also have exponential profiles that extend over many scale lengths (e.g., \citealt{he13}). These galaxies have rising rotation curve potentials, generally do not have spiral waves, and frequently have vigorous star formation in relatively massive clumps accompanied by shells and holes from feedback. Molecular gas discs also have an approximately exponential profile, and \citet{st18} found that fountain flows in star-forming discs can generate these profiles. Stars born out of the molecular discs should inherit the exponential profile as well as low eccentricity orbits. Thus, it appears that a variety of scattering processes generate exponential disc surface density profiles, and can account for the appearance of these profiles in a wide range of galaxies.

The questions remain of how scattering generates these universal profiles, and why the exponential form is preferred? We have addressed these questions from two different directions in recent work. Firstly, in \citet{el16}, we used stochastic scattering models to show that exponential profiles are produced in a disc when there is a slight inward bias in the scattering probability. As is well known, unbiased scattering in one dimension produces a Gaussian distribution. We also showed that such a bias is indeed present in clump scattering models. This stochastic scattering result provides a fundamental kinetic basis for the exponential profile.

Secondly, we examined hydrostatic solutions to the Jeans equations. When further constrained to yield minimum radial entropy gradients, the surface density solutions of these equations have the form of power-laws times Sersic or Einasto profiles. The \citet{se63} form was derived for fitting galaxy surface brightness profiles, and the \citet{ei65} form is a proposed radial density function for early-type galaxies, but both have the form of exponential functions of a radial power-law. Recently, so-called cored S\'{e}rsic profiles, with a polynomial multiplying a S\'{e}rsic function have also been used for bulge and elliptical profile fitting (\citealt{gr03}, \citealt{tr04}, \citealt{sa16}). In fact, the density distribution in the biased scattering models also had the form of an $1/r$ term times the exponential.

Over an interesting parameter range these power-S\'{e}rsic profiles (as we will refer to them) have relatively little curvature in graphs of the logarithm of the surface density versus radius. As a result they can generally be well fit by a simple exponential or with two exponentials for inner and outer radii. This means that they can fit observed profiles in either instances of a Type I profile (simple exponential) or Type II, III profiles (double exponentials with either a downturn or an upturn at large radii).

Recently,  \citet{he17} presented analytical models which are rather similar to those of \citet{st17a}. In these models the stellar orbits in discs with fixed halo potentials were assumed to be mixed by radial migration in a manner that conserved total mass and angular momentum while maximized the net entropy. (The models of \citet{el13} and \citet{st17a} do not conserve stellar angular momentum, but rather exchange it with gaseous scattering centers.) The extremal entropy constraint yields the result that the distribution of stellar angular momentum is always exponential regardless of the potential in such models. These models produce surface density profiles much like those of \citet{st17a} and equation \eqref{eq3} below, though with some additional dependences on the form of the rotation curve. In the case of a flat rotation curve, the \citet{he17} profile is the same as equation \eqref{eq3} with $p = 1$. The authors found that the analytic models compared well with simulations of evolving discs with external perturbations or added internal scattering centers.

It is reassuring that all the various models, including stochastic scattering and Jeans equation models, yield the same type of solutions, i.e., power-S\'{e}rsic. And it is interesting, that in neither case do they generally yield the pure exponential generally used to fit disc profiles, though again we emphasize the difference is not generally large except in the innermost regions. The exception to this generalization is at small radii, well within the scale radius of the exponential or S\'{e}rsic function. While those functions flatten at small radii, the power-law term dominates in the power-S\'{e}rsic profiles. However, at small radii this deviation is likely to be attributed to separate bulge or nuclear components.

Although the biased scattering and Jeans models both yield power-S\'{e}rsic solutions, the former yields essentially one such solution, and the latter allows a wide range of power-law and S\'{e}rsic terms. The reason for this difference is not obvious. Biasing the scattering in different ways would yield a broader range of solutions. \citet{el16} found that a different bias at small and large radii could yield a double exponential profile, so different radially dependent bias factors could probably generate a range of profiles. Alternately, it could be that the Jeans and minimum entropy equations do not provide sufficient constraints for hydrostatic discs. There do not appear to be any additional constraints in discs, or the outer parts of discs, dominated by the halo potential, like those considered in \citet{st17a}.

In self-gravitating discs (see Sec. 2), or partially self-gravitating discs (Sec. 3), an additional constraint is given by the Poisson equation. A primary goal of this paper is to extend the hydrostatic results of \citet{st17a} to such cases, and also include hydrostatic equilibrium in the vertical direction. We will see that the additional constraints do indeed limit the range of the hydrostatic solutions, and lead to a preference for the $e^{-(r/a + z/h)}/r$ form.

\section{Self-gravitating hydrostatic profiles}

In this section we study fully self-gravitating steady state disc profiles determined by the time independent Jeans' equations of stellar hydrodynamics and the Poisson equation. We will neglect azimuthal variations, and assume that net radial velocities are zero across the disc. Then all terms of the mass continuity equation are zero, we have the radial hydrostatic equation in all cases, and a vertical hydrostatic equation in three-dimensional cases with cylindrical symmetry. In the following subsections we consider first two-dimensional systems, then three-dimensional discs, and finally, the constraints imposed by minimizing the entropy gradient in discs.

\subsection{Self-gravitating, two-dimensional S\'{e}rsic discs}

We begin in this subsection by considering a family of two-dimensional disc models. We  adopt the following specific form for the derivative of the gravitational potential,

\begin{equation}
\label{eq1}
\frac{d\Phi}{dr} = \frac{\Phi_o}{r} e^{- \frac{b}{p} \left( \frac{r}{a} \right)^p},
\end{equation}

\noindent where $r$ is the radius in the disc, $\Phi_o$ is a constant potential, $a$ is a scaling factor, $b$ is a constant, and the exponent $p$ is the inverse of the usual S\'{e}rsic index. This particular power-S\'{e}rsic gravitational acceleration is chosen because it is the simplest form that produces the desired general power-S\'{e}rsic surface density profile (see below). We note that that the potential corresponding to this gravity is generally an incomplete gamma function. Gamma function potentials corresponding to S\'{e}rsic-Einasto (or generalized de Vaucouleurs) profiles in spheroidal galaxies have been studied for some time, see e.g., \citet{ciotti91} and \citet{cardone05}. \citet{terzic05} have also studied the incomplete gamma function potentials corresponding to cored-S\'{e}rsic profiles in spheroidal galaxies. For positive values of their argument, here scaled radius, incomplete gamma functions are smooth, and they are reasonable forms for the physical variables, though not as easy to work with as power-law S\'{e}rsic functions. Given the power-S\'{e}rsic hydrostatic disc profiles from \citet{st17a}, equation \eqref{eq1} and the corresponding potential are a natural extension of these works.

Assuming the halo potential is negligible in the present case, the cylindrically symmetric Poisson equation is,

\begin{equation}
\label{eq2}
\frac{1}{r} \frac{\partial}{\partial r} r \frac{\partial \Phi}{\partial r}
+ \frac{\partial^2 \Phi}{\partial z^2}
= -4 \pi G \Sigma,
\end{equation}

\noindent where $\Sigma$ is the disc surface density. In this subsection we neglect the z-derivative term and consider $\Phi$ to be the integral of the potential energy over $z$; $\Phi_0$ has units of velocity-squared times length. Substituting equation (1), and solving for $\Sigma$, we obtain,

\begin{equation}
\label{eq3}
\Sigma (r) = \frac{b \Phi_o}{4 \pi G a^2} \left( \frac{r}{a} \right)^{p-2}
e^{- \frac{b}{p} \left( \frac{r}{a} \right)^p},
\end{equation}

\noindent Confirming that the surface density is a single power-law times a S\'{e}rsic exponential, with an index related to the S\'{e}rsic index. It is interesting that the $p = 2$ case gives a Gaussian surface density, and a near Gaussian gravity. The choice $p = 1$ gives a near exponential surface density, but with the $1/r$ term as well. The gravity has the same form in this case.

The key to the simple solutions is the $1/r$ term in the gravitational acceleration, which cancels the $r$ term between the two derivatives of the radial part of the Laplacian. Using any different power-law term in equation \eqref{eq1} would yield a sum of power-S\'{e}rsic terms for the surface density. Such solutions would generally introduce a two part density and, according to the results of Sec. 2.3 below, a two-part velocity dispersion profile. The resulting (stellar) pressure profile would be even more complex. The velocity dispersion is produced by scattering and, unless there is a change in the nature of the scattering process with radius, these alternate solutions are unphysical. Thus, the family of potential-density pairs given by equations \eqref{eq1} and \eqref{eq3} yield the only monotonic power-law, exponential or power S\'{e}rsic solutions that are physically relevant for two-dimensional scattering discs.

Next we consider the steady radial momentum equation for the stellar disc, which can be written,

\begin{equation}
\label{eq4}
\frac{d\Phi}{dr} - \frac{v_{\theta}^2}{r} = \frac{-1}{\Sigma} \frac{\partial}{\partial r}
\left( \Sigma \sigma ^2 \right) = - \frac{\partial \sigma ^2}{\partial r}
- \frac{ \sigma ^2}{\Sigma} \frac{\partial \Sigma}{\partial r},
\end{equation}

\noindent where $v_{\theta}$ is the local mean azimuthal velocity, and $\sigma (r)$ is the radial velocity dispersion at radius $r$. As discussed in \citet{st17a} there is no reason to assume that the three terms in the first equality: gravity, centrifugal and pressure gradient accelerations, all scale in the same way with radius. We will follow that paper in adopting a centrifugal imbalance function, $\chi(r)$, to describe the different scalings. Specifically, $\chi$ is defined as the fraction of the gravity balanced by the pressure gradient, or equivalently, as the `centrifugal imbalance,'

\begin{equation}
\label{eq5}
\frac{v_{\theta}^2}{r} = \left(1-\chi(r) \right)  \frac{d\Phi}{dr}.
\end{equation}

With this definition, we can then write the pressure part of the equation \eqref{eq4} as,

\begin{equation}
\label{eq6}
\chi \frac{d\Phi}{dr} \Sigma = -\frac{\partial}{\partial r}
\left( \Sigma \sigma ^2 \right).
\end{equation}

\noindent Following substitution for the gravity and surface density from equations \eqref{eq1} and \eqref{eq3} this becomes an equation for two unknowns: $\chi$ and $\sigma^2$. To balance this equation at least one of these variables must contain a S\'{e}rsic-type exponential term. If it is the $\chi$ variable, then the exponential term must be increasing with radius, unless the $\sigma^2$ term is even more strongly increasing. Either alternative is unphysical in the context of galaxy discs. The simplest expedient is to assume that $\sigma^2$ has the same exponential term as the gravity and the surface density. Then there is no need for an exponential factor in $\chi$, yielding the more palatable result that it varies more slowly, in general, than the other variables.

Specifically, we adopt a power-law S\'{e}rsic form for the velocity dispersion, like that of the gravity and the surface density, but with an unknown power $n$,

\begin{equation}
\label{eq7}
\sigma^2 = \sigma_o^2 \left( \frac{r}{r_{\sigma}} \right)^n
e^{- \frac{b}{p} \left( \frac{r}{a} \right)^p},
\end{equation}

\noindent where $\sigma_o$ is a constant, and the scale length $r_{\sigma}$ is not assumed to equal $a$. Admittedly, this form is an ad hoc choice at this point. The entropy gradient constraints of Sec. 2.3 provide a specific motivation for choosing it.

Then we can solve equation \eqref{eq6} for $\chi(r)$, to obtain,

\begin{equation}
\label{eq8}
\chi(r) = \left( \frac{a}{r_{\sigma}} \right)^n
\left[ (2-n-p) \left( \frac{r}{a} \right)^n + 2b \left( \frac{r}{a} \right)^{p+n} \right]
\frac{\sigma_o^2}{\Phi_o}.
\end{equation}

\noindent Physically reasonable values of $\chi$ are positive, so we expect $n + p$ not to be very large and to be positive. To avoid exponential terms that grow with radius, $p$ must be positive.

An especially interesting family of models are those with $n = -p$. Then,

\begin{equation}
\label{eq9}
\chi(r) = 2 \left( \frac{a}{r_{\sigma}} \right)^{-p}
\left[ \left( \frac{r}{a} \right)^{-p} + b  \right]
\frac{\sigma_o^2}{\Phi_o},
\end{equation}

\noindent which goes to a constant value at large values of $r/a$ (and which must be less than 1.0 according to equation \eqref{eq5}). In a self-gravitating, flat rotation curve disc we might expect $p$ to be small to obtain a gravitational acceleration with an approximately $1/r$ fall-off. Then $\chi$ and $\sigma$ would also vary slowly with radius. However, in that case, $\Sigma$ has power-law form; a self-gravitating exponential disc requires a halo to produce a flat rotation curve.

In cases where $n > p$ the power-law term in equation \eqref{eq7} can partially balance the exponential over some range of radii, yielding a more slowly varying dispersion. However, in this case, the centrifugal imbalance would grow rapidly with radius unless both $n$ and $p$ are small.

Despite the fact that models given by equations \eqref{eq1} - \eqref{eq8} are very simple, they can produce a significant range of models as the parameters are varied. Moreover, they can be readily generalized to much broader model families, e.g., by adding more power-law exponential terms to the right-hand-side of equation \eqref{eq1}. Equations \eqref{eq3} and \eqref{eq7} would have corresponding terms added. However, because of the nonlinear terms in the radial hydrostatic equation \eqref{eq4}, the centrifugal imbalance equation (8) would become considerably more complicated. If, as assumed, the steady state is mediated by scattering processes, we would expect smooth, monotonic forms for all of these variables, and thus, a bias towards the simplest forms.

Another important point concerns the fact that while discs may be largely self-gravitating in their inner parts, gravity in their outer parts is generally halo dominated. The fixed potential solutions of \citet{st17a} have the same form of the surface density as equation \eqref{eq3}, though the dispersion profile is of power-law form (depending on the form of the halo potential). Thus, there could be a very smooth transition from one regime to the other, a point we will take up later.

\subsection {Three-dimensional systems with cylindrical symmetry}

The two dimensional disc solution above can be generalized to three dimensional systems with cylindrical symmetry, specifically, cylindrical rotation. Such solutions are relevant to thick discs and pseudo-bulges. There are a number of possible generalizations of the formulae above, but the one that yields the simplest profile forms begins with a potential gradient of the form,

\begin{equation}
\label{eq13}
\nabla \Phi = \frac{\Phi_o}{r} e^{- \frac{b}{p} \left( \frac{r}{a} + \frac{z}{h} \right)^p}
\left( 1, c \right).
\end{equation}

\noindent Note the additional exponential term for the vertical direction $z$ with scale factor $h$. The vector $(1, c)$ gives amplitudes for the radial and vertical directions, with $c$ assumed to be a positive constant.  Secondly, we note that any potential with multiplicative $r, z$ terms will couple the corresponding accelerations. E.g., with the above form the radial acceleration will decrease with scale height. However, where this does not correspond to observation, the effect can be minimized with a substantial value of the scale height, $h$. This also highlights the fact that different potentials are needed for different disc components, but we will only consider one component here.

Then the Poisson equation, with $\Phi_0$ now having the units of potential, gives a density of

\begin{multline}
\label{eq14}
\rho (r, z) = \frac{b\Phi_o}{4 \pi G a^2}
\left( 1 + \frac{ac}{h} \right)
\left( \frac{r}{a} \right)^{-1}\\
\times \left( \frac{r}{a} + \frac{z}{h} \right)^{p-1}
e^{- \frac{b}{p} \left( \frac{r}{a} + \frac{z}{h} \right)^p}.
\end{multline}

\noindent This form may not look particularly simple, but with $c, z = 0$ it is the same as equation \eqref{eq3}. Any other form of the z-dependence in the potential gradient yields more complex forms for the density, and infinite series or special function forms for the dispersion.

Next we substitute the radial term of the potential gradient and the density into the radial momentum equation \eqref{eq4}. We further assume that the radial component of the velocity dispersion has the form (we will return to this point shortly),

\begin{equation}
\label{eq15}
\sigma_r^2 = \sigma_o^2 \left( \frac{r}{r_\sigma} \right)^{j}
e^{- \frac{b}{p} \left( \frac{r}{a} + \frac{z}{h} \right)^p}.
\end{equation}

\noindent Then the radial momentum equation gives the following form for the centrifugal imbalance function,

\begin{multline}
\label{eq16}
\chi(r) = \frac{\sigma_o^2}{\Phi_o} \left( \frac{r}{r_{\sigma}} \right)^j
\left( \frac{r}{a} + \frac{z}{h} \right)^{p-1}  \times\\
\left[ 1-j - (p-1) \frac{\frac{r}{a}}{\frac{r}{a} + \frac{z}{h}}
+ 2b \frac{\frac{r}{a}}{\left( \frac{r}{a} + \frac{z}{h} \right)^{1-p}} \right].
\end{multline}

\noindent The various parameters in this and the preceding expressions allow for a wide range of models, despite their relative simplicity. Before continuing, we note the special case with $p = 1, j = -1$, which has density and radial velocity profiles of $(1/r)$ times an exponential form. Its centrifugal imbalance function is,

\begin{equation}
\label{eq17}
\chi(r) = 2\frac{\sigma_o^2}{\Phi_o}  \left[ \frac{a}{r} + b \right].
\end{equation}

\noindent This centrifugal imbalance is large at small radii (little rotational support), but goes to  a constant at large radii. The radial dispersion and density are also large at small radii. Thus, this case could represent both a secular bulge at small radii and a disc at large radii, if the constant $\chi$ is small there.

Next we consider the vertical momentum equation, which balances the vertical pressure gradient against the gravity (with no centrifugal term). This equation can be written,

\begin{equation}
\label{eq18}
\frac{\partial \Phi}{\partial z} \rho = -\frac{\partial}{\partial z}
\left( \rho \sigma_z^2 \right).
\end{equation}

\noindent We adopt a form for the vertical velocity dispersion like that of equation \eqref{eq15}, i.e.,

\begin{equation}
\label{eq19}
\sigma_z^2 = \sigma_{zo}^2 \left( \frac{r}{r_\sigma} \right)^{-1}
\left( \frac{r}{a} + \frac{z}{h} \right)^{1-p}
e^{- \frac{b}{p} \left( \frac{r}{a} + \frac{z}{h} \right)^p}.
\end{equation}

\noindent Then substituting this and the above expressions for the density and the gravity into equation \eqref{eq18}  yields an equation in three different powers ($p-1, 2p-3$, and $3p-3$) of the term $(r/a + z/h)$. Generally, this is not solvable for any arbitrary value of $p$, and we should seek a more general (infinite series) form for $\sigma_z^2$. As per the discussion of the previous subsection, the presence of a single scattering process suggests that a multi-part velocity dispersion is unphysical. However, in the special case of $p = 1$, two of these term powers disappear, while the third has a coefficient of $p-1$, so it also disappears. The equation reduces to a relation between coefficients,

\begin{equation}
\label{eq20}
\sigma_{zo}^2 = \frac{ch \Phi_o}{2ab},
\end{equation}

\noindent and the vertical velocity dispersion is also an exponential in $z$ (as well as retaining the $1/r$ dependence).

This result realizes the goal stated at the end of the introduction, that the inclusion of disc self-gravity would narrow the large family of power-S\'{e}rsic profile forms found in \citet{st17a}. Moreover, the preferred solution here, $(1/r)$ times an exponential, is the same as that found in biased scattering models in ES16. However, the solutions above are based on some key assumptions about the form of the velocity dispersions. In the following subsection we provide further justification of those assumptions on the basis of minimal entropy gradients.

Another property of solutions with accelerations of the form of equation \eqref{eq13} is that their rotation curves decline exponentially with $r/a$ and $z/h$. Thus, while it can have a relatively flat form for $r \ll a$, a halo potential is needed to retain a flat form from $r \approx a$, and outward. These solutions might account for the fact that after an initial rise in the (bulge or core-dominated) innermost regions, rotation curves often decline from a peak before flattening out in discs.

\subsection{Zero Entropy Profile Constraint}

In \citet{st17a} we argued that, even in the presence of some long-range scattering, the basic thermodynamical relations should still hold, at least approximately. This includes the fundamental relation of thermodynamics, which we wrote as,

\begin{equation}
\label{eq21a}
T \frac{dS}{dr} = \frac{dE}{dr} + P\frac{d}{dr} \left( \frac{1}{\rho} \right)
+ \frac{d\alpha}{dr},
\end{equation}

\noindent where $S$ is the entropy, $E$ the internal energy per unit mass, $P$ the pressure, $\rho$ the density and $T$ is the temperature (e.g., a multiple of the mean dispersion squared). We also include a free energy term, $\alpha$, which is the energy gain or loss from stars scattered in or out of any local volume, see \citet{st17a}. As in that paper, we have not included a gravitational potential gradient term on the assumption that the gradients above are determined primarily by local scattering, that is, local relative to the potential scale length.

We further assume that $E = c_E \sigma^2, P = c_E \rho \sigma^2$, and $kT = c_T \sigma^2$. Here $\sigma^2$ represents an appropriate mean velocity dispersion, and the constant coefficients $c_E, c_T$ have values appropriate to the given case, i.e., two or three dimensional discs, with radial or radial and azimuthal dispersions included. After the substitution of these forms into equation \eqref{eq21a}, that equation can be reduced to the following form in the limit of zero entropy gradient ($dS/dr = 0$),

\begin{equation}
\label{eq21b}
\frac{1}{\sigma^2} \frac{d\alpha}{dr} =
 \frac{d}{dr}  \left(ln \frac{\rho}{\sigma^2} \right).
\end{equation}

When the gradient of $\alpha$ is small (little variation in the scattering with radius) or the velocity dispersion is relatively large, then the right hand side of this equation is also small, and we have $\rho (r) \sim\sigma^2(r)$.

This last result is consistent with the forms we adopted for the radial and vertical velocity dispersions in the previous subsection. Now we can view the relatively simple models that result with those forms of the dispersion as a consequence of minimal entropy gradients and uniform scattering within the disc. This, in turn, provides a reason for the universality of those forms discussed in the Introduction.

Viewing these results from a somewhat different point of view, with negligible entropy and $\alpha$ gradients, we are left with the middle two terms in equation \eqref{eq21a}. The remaining equation is somewhat like an adiabatic law, where gas cools as the density decreases. This analogy suggests the pressure can be expressed as $P \propto \rho^{\gamma}$. Consistency with $\rho \propto \sigma^2$ yields $\gamma = 2$. {\it The constancy of $\rho \propto \sigma^2$ also gives a constant scale height, by the usual definition $h$, and may help explain that somewhat mysterious result observed in many cases} \citep[e.g.,][]{vdk82, va02}.

\section{A family of self-gravitating disc plus fixed halo models}
\subsection{The model family}

Discs are not isolated, self-gravitating structures. The dark halo potential dominates in a large, outer part of a disc (assuming conventional gravity theory). The solutions of the previous section are most applicable to the inner disc, while those of \citet{st17a} apply to the halo-dominated regions. Of course, both potentials play a significant role in intermediate regions. In this section we present a sample family of solutions in the case of the gravity consisting of a sum of both potentials. We begin by assuming a potential gradient of the following form for the disc plus halo,

\begin{equation}
\label{eq22}
\nabla \Phi = \frac{\Phi_o}{r} e^{- \frac{b}{p} \left( \frac{r}{a} + \frac{z}{h} \right)^p}
\left( 1, c \right) + \frac{\Phi_1}{r} \left( 1, 0 \right).
\end{equation}

\noindent This halo potential is clearly a special case, chosen here for its simplicity; it is the logarithmic potential which gives the common flat rotation curve. The Poisson equation is linear in both the potential and density terms, and so, disc and halo parts are separable, i.e., essentially separate Poisson equations.

We further assume a thin disc here, so that the halo density within the disc and the contribution of the halo potential to the vertical force are negligible. This approximation suggests that the vertical density profile can still be described by equation \eqref{eq14}, even beyond the radius where the radial gravity is dominated by the halo contribution. Since the radial part of the profile is essentially the same as that used with halo potentials in \citet{st17b}, we also expect that that form could be extended to the outer disc.

Let us reconsider the radial hydrostatic equation, using this approximation that the stellar density is unchanged between the present model and that of the previous subsection. The only variable with a nonlinear dependence in this hydrostatic equation is the density, so we can separate the terms associated with the halo potential, i.e., in the gravity and dispersion terms, and treat them separately as in the Poisson equation.  We add the following term to the radial dispersion of equation \eqref{eq15},

\begin{equation}
\label{eq23}
\sigma_{rx}^2 = \sigma_{1}^2 \left( \frac{r}{r_x} \right)^{k}.
\end{equation}

The halo part of the radial momentum equation becomes,

\begin{equation}
\label{eq24}
\chi \frac{\Phi_1}{r} \rho = -\frac{\partial}{\partial r}
\left( \rho \sigma_{rx} ^2 \right).
\end{equation}

After substitution of the expressions above for the halo gravity, radial velocity dispersion, and the centrifugal imbalance function of equation \eqref{eq16} this reduces to a scaling between constants of the disc/halo components,

\begin{equation}
\label{eq25}
\frac{\sigma_{1}^2}{\sigma_o^2} = \frac{\Phi_1}{\Phi_o}
\left( \frac{r_{\sigma}}{r_x} \right) ^k.
\end{equation}

\noindent In the special case where $r_{\sigma} = r_x = a$ we have the very simple relation $\frac{\sigma_{1}^2}{\sigma_o^2} = \frac{\Phi_1}{\Phi_o}$.

As a result of the assumption of a very thin disc containing negligible halo mass, the $z$ hydrostatic equation is unchanged from the previous subsection, with no new component in the vertical velocity dispersion.

\subsection{Entropy gradients revisited}

On the face of it, equation \eqref{eq24} is rather strange. It relates the centrifugally reduced halo gravity to an extra component of the stellar radial velocity dispersion. Why should the centrifugal balance function continue smoothly from the disc gravity dominated region to the region of halo dominance? How does the two-part radial velocity dispersion (with terms from equations \eqref{eq15} and \eqref{eq23}) allow the radial entropy gradient to continue to be zero?

We believe that the answer to both questions derives from the assumption that scattering remains smooth across both parts of the disc. This is reasonable if the scattering is due to gaseous clumps, large scale waves or external perturbations. To maintain a zero entropy gradient in the halo dominated region, equation \eqref{eq21b} must still be satisfied. However, the relation $\rho (r) \sim\sigma^2(r)$ is no longer true. The density is still of the form of equation \eqref{eq14}, but the total velocity dispersion term is the sum of the two radial dispersion terms and the z-term from equation \eqref{eq19}. Generally, the dispersion squared - density ratio will have strong radial variations, and then, so too will either the free energy gradient or the entropy gradient.

There is one special case of note, when $p = 1$. In that case, power-law terms disappear or cancel in the $\rho/\sigma^2$ ratio, and it reduces to the form $c_1/(1 + c_2$ exp(r/a)) (with constants $c_1,\ c_2$). This further reduces to a constant, as above, for $r \ll a$, or to the term $(c_1/c_2)$exp(-r/a) when $r \gg a$. In the latter case, according to equation \eqref{eq21b}, the free energy gradient is a constant, and the entropy gradient zero. Indeed, the constants depend on the dispersion scales (e.g., $\sigma_o,\ \sigma_1,\ \sigma_{z1}$), or the density scale, so with appropriate values of these, the free energy gradient can be made to vanish, as it does at small radii. This is how an entropy gradient, globally minimized by smooth scattering, can account for the equations of the previous subsection and a well regulated stellar disc. We will explore the $p = 1$ model with a detailed example in the following section.

\subsection{Azimuthal velocities}

Up to this point we have not explicitly considered azimuthal velocities and rotation curves. However, the present model contains the preferred model of the previous section as a limiting case, so it is efficient to do it in the context of this present model. According to the definition of the centrifugal imbalance, the azimuthal velocities are given by,

\begin{equation}
\label{eq26}
\frac{v_{\theta}^2}{r} = \left[ 1 - \chi(r) \right] \nabla \Phi.
\end{equation}

\noindent Consider the two limits of this expression. First, at large radii ($r/a \gg 1$), $\chi$ goes to a constant value, and the gravity goes as,

\begin{equation}
\label{eq27}
 \nabla \Phi \sim \frac{1}{r} \left( \Phi_o
 e^{- \frac{b}{p} \left( \frac{r}{a} + \frac{z}{h} \right)}
 + \Phi_1 \right) \sim \frac{1}{r},
\end{equation}

\noindent then $v_\theta$ is constant, and the rotation curve is flat, as expected with the adopted halo potential.

The second limit is at small $r/a$, where $\chi \sim a/r$. In this limit, as $r$ gets smaller, $\chi$ grows, the azimuthal velocity decreases inward, and we have a rising rotation curve. At the same time, $\sigma_z, \sigma_r \sim 1/r^{1/2}$, so the dispersions grow as radius decreases. Thus, the innermost regions might be better described as a secular bulge in this model, and one that is a natural extension of the disc solution. Specific examples of rotation curves will be given in the next section.

\subsection{Disc masses}

We briefly note in this subsection that the preferred ($p=1$) model of this and the previous section has a very simple formula for its total mass, as in the case of the pure exponential. The mass integral is,

\begin{equation}
\label{eq28}
 M(r_{max}) = 2\pi \int_0^{r_{max}} r dr
 \int_0^\infty 2 \rho (r,z) dz,
\end{equation}

\noindent where $r_{max}$ is the outer radius of the disc, and the factor of 2 in the inner integral accounts for regions above and below the mid-plane. If we substitute the density from equation \eqref{eq14}, and take $b = p=1$ (with no loss of generality), then both integrals can be evaluated explicitly and we obtain,

\begin{equation}
\label{eq29}
 M(r_{max}) = \frac{\Phi_o}{2G}
\left( h + ac \right)
\left[ 1 - e^{-r_{max}/a} \right]
\simeq \frac{\Phi_o}{2G}
\left( h + ac \right),
\end{equation}

\noindent which depends primarily on the disc scale lengths ($a, h$) and potential amplitude ($c$), and only weakly on disc size for $r_{max} \gg a$. Alternately, the model suggests a relation between disc scale length and disc mass.

We note that if we set $r_{max} = a$ in equation \eqref{eq29}, assume $ac >> h$, and that $\Phi_o \propto a$, then $M(a) \propto a^2$, and the central surface brightness within a radial scale height will be constant \citep{fr70}. The proportionality $\Phi_o \propto a$ must be the result of evolutionary processes not considered in this work.

\subsection{Scale heights and outer disc flaring}

Equation \eqref{eq19} suggests that the vertical velocity dispersion in these models falls rapidly with increasing radius, though the scale height remains constant as observed in many edge-on galaxies. At the same time, discs are often seen to flare at large radii \citep[e.g.,][]{sarkar19}. There are several possible causes of this discrepancy. One is simply the neglect of vertical gravity contributed by the halo potential. This approximation should be good as long as the outer disc remains thin. However, external disturbances can pull stars out of thin discs. Then the vertical part of the halo gravity becomes significant, and the flared solutions are applicable.

\section{Simple disc examples}

Considering the number of parameters involved, the models of the previous section make up a large family. We will not attempt to thoroughly explore that family, but in this section we will present a few particular examples to demonstrate the plausibility of these models. Given the discussion of smoothness and minimum entropy gradients in Sec. 3.2, we confine our attention to the smooth $p = 1$ models i.e., pure exponential density profile with a $1/r$ factor. (Note that such functions are called exponential polynomials in mathematics.) 

To further simplify we take: $b = 1$ in the exponential factors, $c = 1$ in equation \eqref{eq22}, $j = -1$ in equation \eqref{eq15}, and $k = -1$ in equations \eqref{eq23} and \eqref{eq25}. The latter two choices are consistent with zero free energy and entropy gradients, see above. As discussed below equation \eqref{eq25} we will also assume that the constants satisfy $r_{\sigma} = r_x = a$. 

We adopt specific parameter values that are like those of the solar neighborhood, see \citet{bt08}. Specifically, we take a representative radius of $r_o = 10\ kpc$, an exponential scale length of $a = 2.5\ kpc$, a vertical scale height of $h = 300\ pc$, and a stellar density at $r_o$ of $\rho_o = 0.6 M_{\odot}pc^{-3}$. Additionally we adopt a radial velocity dispersion at $r_o$ of $\sigma_r(r_o) = 40\ km\ s^{-1}$, and vertical dispersion at $r_o$ of $\sigma_z(r_o) = 20\ km\ s^{-1}$. Then we have - $a/h = 8.33$, and $a/r_o = 0.25$.

With these values the density equation (eq. \eqref{eq14}) reduces to $\Phi_o = 294G \rho_o a^2 = 4.73 \times 10^{11}\ m^2s^{-2}$. The total radial velocity dispersion is given by the sum of equations \eqref{eq15} and \eqref{eq23}. The ratio of the two scale dispersions is given in terms of the potential constants by equation \eqref{eq25}, i.e., $\sigma_1^2 /\sigma_o^2 = \Phi_1/\Phi_o$. The latter ratio is determined by the radius were the gravitational acceleration of the halo first exceeds that of the disc. We will assume that the value of that radius is $r_{eq} = 2a = 5\ kpc$. Then we obtain $\Phi_1/\Phi_o = 0.135$. Then, using the value of the ratio of $\sigma_1 /\sigma_o = 0.367$, and the value of $\sigma_{ro}$, the radial dispersion equation yields, $\sigma_o = 204\ km\ s^{-1}$. Similarly, using the value of the vertical dispersion at $r = r_o$, the vertical dispersion equation (eq. \eqref{eq19}) yields $\sigma_{zo} = 296\ km\ s^{-1}$.

Next we can use the values for $\Phi_o, \Phi_1$ in equation \eqref{eq22} to get the gravitational acceleration in this case. This can then be substituted into equation \eqref{eq5}, along with equation \eqref{eq17} which gives the appropriate form of the centrifugal imbalance function $\chi$ in this case, to get the azimuthal velocity. Specifically, in this case (with $z = 0$) we obtain,

\begin{equation}
\label{eq30}
v_{\theta}^2 = \left( 1-\chi \right) r \frac{d\Phi}{dr}
=  {\Phi_o} \left[ 1 - \frac{2{\sigma_o^2}}{\Phi_o}
\left(\frac{a}{r} + 1 \right) \right]
\left[ e^{-r/a} + \frac{\Phi_1}{\Phi_o} \right].
\end{equation}

\noindent With the particular values of the constants above we find, $v_{\theta}(r_o) = 238\ km\ s^{-1}$.

The radial variations of the mid-plane density, azimuthal velocity and the velocity dispersions of this example model are shown in Fig. 1. As expected the density is exponential except in the innermost region, where the $1/r$ factor produces an upturn. The outer rotation curve is flat, but there is a rather large bump within one scale length. As we will show below this is a result of the particular parameter choices, not an essential feature of the model. The dispersions slowly decrease in the outer regions, but in the inner regions they increase rapidly. The onset radius of this phenomenon could be pushed inward by adjusting the values of the constants $r_x, r_{\sigma},$ and $b$, or in the case of the azimuthal velocity by adjusting $r_{eq}$. (Recall that this factor was set to $2a$, which explains why $v_{\theta}$ rises at that radius.) However, the cusps could not be eliminated by this means.

Velocity dispersion profiles without cusps are likely a kinetic phenomenon, not captured by hydrostatic models. That is, the radial excursion of stellar orbits or the scattering distance exceeds the mean radius, and then the density and velocity profiles flatten out or decrease. Indeed, the tendency towards high velocities and dispersions in the inner regions would generate larger radial excursions, central smoothing, and perhaps, even bulge formation. This process is driven by scattering, {\bf but} another possibility is that the scattering is not strong enough to produce the high velocities of the hydrostatic model. In that case, the relaxation to the hydrostatic form {\bf may never be} achieved in the innermost disc. In the outer regions the profiles are quite realistic, and satisfy the various physical constraints, including the minimum entropy gradient constraint.

\begin{figure}
\centerline{
\includegraphics[scale=0.36]{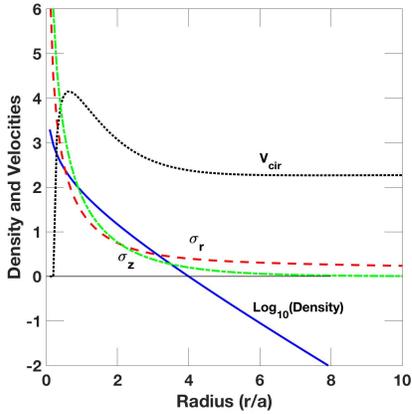}}
\caption{Density and velocities profiles in the example model. Solid blue curve gives the logarithm of the mid-plane density in units of $0.6\ M_{\odot}\ pc^{-3}$, the assumed value at $r/a = 4$. The black dotted curve gives the circular velocity, the red dashed curve gives the radial velocity dispersion, and the green dot-dash curve gives the vertical velocity dispersion, all in units of $100\ km\ s^{-1}$ for comparison.}
\end{figure}

More examples of model rotation curves are provided by Fig. 2. The only difference between the curves in the figure is in the value of the parameter $\Phi_o$. Note, we keep the ratio of $\Phi_1/\Phi_o$ constant, so the halo gravity changes with the disc self-gravity. Then there {\bf are} two effects of decreasing $\Phi_o$ according to equation \eqref{eq30}. First, the velocities at inner radii are decreased, and second, the centrifugal imbalance factor is increased substantially. The more $\Phi_o$ is decreased, the more the centrifugal imbalance drives the rotation curve down before the bump appears. The disc density is also proportional to $\Phi_o$, so this can also be seen as the effect of a lower density disc given a fixed velocity dispersion. The range of these rotation curve forms largely encompasses that observed (see e.g., \citealt{so16}), even though we have not yet investigated the full range of model possibilities. The prediction that the curve shapes in Fig. 2 near $r = a$ depend on the disc-to-halo mass ratio (or halo compactness) may be testable. It is true that rising rotation curves are commonly found in dwarf galaxies, where the halo may be less compact relative to the disc size (e.g., \citealt{sw09}).

What is the nature of the bump in some of these rotation curves? The innermost rising part of the rotation is due to the centrifugal imbalance term in these models, e.g, a hotter inner disc. If the disc gravity continues to dominate over the halo at or beyond $r \approx a$, then the exponential term will drive the circular velocity down rapidly with increasing radius, until the halo dominates, and the rotation curve flattens. The bump can be taken as evidence for the lack of a disc-halo conspiracy in those galaxies that show the feature. In galaxies with rotation curves like the lower two in Fig. 2, there is also no conspiracy because disc self-gravity is not important over any significant radial range. The reason for an apparent conspiracy in the density and dispersion profiles of Fig. 1 could be that both of those quantities are determined by scattering (and entropy gradient smoothing) in the stellar disc.

\begin{figure}
\centerline{
\includegraphics[scale=0.36]{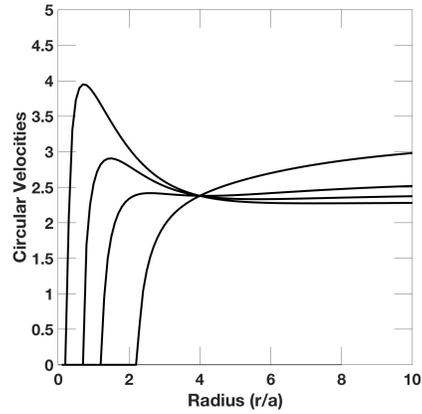}}
\caption{Various rotation curves according to equation \eqref{eq30} with different values of the potential scale $\Phi_o$. The values of this factor for the various curves, from highest to lowest in the inner regions, are $(1.2, 1.5, 2.0, 4.0) \times 10^{11}\ m^2\ s^{-2}$.}
\end{figure}

\section{Summary: universal power-law S\'{e}rsic discs}

In sum, the models of the previous sections provide steady, and physically constrained descriptions of exponential discs with self-gravity in logarithmic halo potentials. As long as conditions like equation \eqref{eq25} are satisfied, the variables are smooth functions of radius. Specifically, unless the potentials have breaks (violating equation \eqref{eq25}), the density and rotation curves should be smooth, though the latter may have a bump feature. In the previous subsection we suggested that the bump may be related to the so-called disc-halo conspiracy (see e.g., \citealt{ke87},  \citealt{be15}). That is, when present the bump may be a feature marking the handoff of disc to halo dominance in the gravity. There is no such feature in the density profile because the same profile form extends across both regions of the disc. The reason for this is that the halo density is negligible within the discs of the smoothest models, and so, does not strongly effect the disc density profile, nor the vertical structure. It only effects the radial dispersion according to equation \eqref{eq23} (but see Sec. 3.5 for a variant model). Similarly, no tuning is required if a secular bulge component hands off to the disc, because in those models they are one and the same profile.

Another key factor that makes the density profile simple is the $1/r$ term in both disc and halo gravities. With a different power, the Poisson equation would give a more complex density or centrifugal imbalance ($\chi$) profile. This term, and the adoption of a pure exponential ($p =-1$), rather than a different S\'{e}rsic term, are also needed to satisfy the zero entropy and free energy constraints in the simple models.

Real galaxy discs may not achieve such well relaxed states, depending on their environment and internal scattering history. This fact, and the possibility of different halo potentials, helps explain why some discs are better fit with other S\'{e}rsic forms, or have profile breaks. The simple models are idealized, but they bring together a wide range of physical constraints, with much symmetry.

The vertical velocity dispersion falls off exponentially with the vertical height $z/h$. The scale height is usually constant, though it may increase in the outer disc. Specifically, in cases when the vertical gravity of the outer disc is dominated by the halo, the radial dispersion does not fall exponentially at large radii, but rather as $r^{1/2}$ with the assumed potential form. The available observations do show a slower fall-off than the surface density, and such a power-law seems a reasonable fit (recent examples include: \citealt{ma13}, \citealt{ge15}, \citealt{mo18} and references therein).

As suggested above the models also have ramifications for the formation of secular bulges. These are thought to for via secular processes out of disc material, and retain disc characteristics, such as metallicity and higher rotation than classical bulges (\citealt{ko04}, \citealt{ko16}). In many cases the formation of secular bulges is thought to be facilitated by the processes of bar dissolution \citep{se14}. Equations \eqref{eq7} and \eqref{eq19} show that in the present models the components of the velocity dispersion increase rapidly toward the disc center. As suggested above, the core surface density probably smooths, e.g., due to kinetic effects, reducing the vertical gravity. Although the velocity dispersions may also be smoothed in the center, it seems likely that scattering (another secular process) will produce a more three-dimensional distribution of stars. Thus, the inner parts of discs may be naturally prone to develop bulge-like features as they relax to hydrostatic forms. This provides an explanation for the development of `disky' versus `boxy' pseudo-bulges (see \citealt{ko16}), i.e., the latter are formed by bar dissolution, and the former by secular disc evolution with scattering.

The solutions above are not mathematically unique, but rather the simplest and smoothest solutions given the various constraints. Solutions with more general power-laws and S\'{e}rsic forms might be relevant in the early stages of disc formation or after global disturbances. However, scattering processes, especially those that generate long-range excursions, drive smoothness. Thus, these processes will tend to favor the simplest and smoothest forms in the long term. Specifically, the universal profile forms of the models are strongly constrained by minimizing the entropy gradient.  These simple models offer many insights into the structure of stellar discs, and they should prove to be very useful tools for understanding galaxy discs.

\section*{Acknowledgments}

We acknowledge use of NASA's Astrophysics Data System, and the NASA Extragalactic Data System.

\bibliographystyle{mn2e}

\bsp
\label{lastpage}
\end{document}